\documentstyle[12pt]{article}
\makeatletter
\def\@sect#1#2#3#4#5#6[#7]#8{\ifnum #2>\c@secnumdepth
    \def\@svsec{}\else
    \refstepcounter{#1}\edef\@svsec{\csname the#1\endcsname.\hskip 1em }\fi
    \@tempskipa #5\relax
    \ifdim \@tempskipa>\z@
    \begingroup #6\relax
    \@hangfrom{\hskip #3\relax\@svsec}{\interlinepenalty \@M #8\par}
    \endgroup
    \csname #1mark\endcsname{#7}\addcontentsline
    {toc}{#1}{\ifnum #2>\c@secnumdepth \else
     \protect\numberline{\csname the#1\endcsname}\fi
           #7}\else
    \def\@svsechd{#6\hskip #3\@svsec #8\csname #1mark\endcsname
          {#7}\addcontentsline
          {toc}{#1}{\ifnum #2>\c@secnumdepth \else
     \protect\numberline{\csname the#1\endcsname}\fi
           #7}}\fi
     \@xsect{#5}}
\def\label#1{\@bsphack\if@filesw {\let\thepage\relax
   \xdef\@gtempa{\write\@auxout{\string
   \newlabel{#1}{{\thesection.\@currentlabel}{\thepage}}}}}\@gtempa
   \if@nobreak \ifvmode\nobreak\fi\fi\fi\@esphack}
\def\@eqnnum{(\thesection.\theequation)}
\def\section{\setcounter{equation}{0} \@startsection {section}{1}{\z@}{-3.5ex
   plus -1ex minus -.2ex}{2.3ex plus .2ex}{\Large\bf}}
\newcount\@minsofar
\newcount\@min
\newcount\@cite@temp
\def\@citex[#1]#2{%
\if@filesw \immediate \write \@auxout {\string \citation {#2}}\fi
\@tempcntb\m@ne \let\@h@ld\relax \def\@citea{}%
\@min\m@ne%
\@cite{%
  \@for \@citeb:=#2\do {\@ifundefined {b@\@citeb}%
    {\@h@ld\@citea\@tempcntb\m@ne{\bf ?}%
    \@warning {Citation `\@citeb ' on page \thepage \space undefined}}%
{\@minsofar\z@ \@for \@scan@cites:=#2\do {%
  \@ifundefined{b@\@scan@cites}%
    {\@cite@temp\m@ne}
    {\@cite@temp\number\csname b@\@scan@cites \endcsname \relax}%
\ifnum\@cite@temp > \@min
    \ifnum\@minsofar = \z@
      \@minsofar\number\@cite@temp
      \edef\@scan@copy{\@scan@cites}\else
    \ifnum\@cite@temp < \@minsofar
      \@minsofar\number\@cite@temp
      \edef\@scan@copy{\@scan@cites}\fi\fi\fi}\@tempcnta\@min
  \ifnum\@minsofar > \z@ 
    \advance\@tempcnta\@ne
    \@min\@minsofar
    \ifnum\@tempcnta=\@minsofar 
      \ifx\@h@ld\relax
        \edef \@h@ld{\@citea\csname b@\@scan@copy\endcsname}%
    \else \edef\@h@ld{\ifmmode{-}\else--\fi\csname b@\@scan@copy\endcsname}%
      \fi
    \else \@h@ld\@citea\csname b@\@scan@copy\endcsname
          \let\@h@ld\relax
  \fi 
\fi}%
\def\@citea{,\penalty\@highpenalty\,}}\@h@ld}{#1}}
\def\appendixname{Appendix}
\def\appendix{\par
  \def\pre@section{\appendixname{} }
  \setcounter{section}{1}
  \@addtoreset{equation}{section}
  \def\thesection{\Alph{section}}
  \def\theequation{\arabic{equation}}}
\makeatother
\begin{document}
\def\t{\theta}
\def\T{\Theta}
\def\w{\omega}
\def\ov{\overline}
\def\a{\alpha}
\def\b{\beta}
\def\g{\gamma}
\def\s{\sigma}
\def\l{\lambda}
\def\wt{\widetilde}
\def\di{\displaystyle}
\def\sn{\mbox{sn}}
\def\cd{\mbox{cd}}
\def\cn{\mbox{cn}}
\def\dn{\mbox{dn}}
\def\tn{\mbox{tn}}

\addtolength{\unitlength}{-0.5\unitlength}

\begin{flushright} July, 1994
\end{flushright}
\vspace{3cm}
\begin{center}
{\bf Modified tetrahedron equations and related 3D integrable models}\\
\vspace{1cm}
{\bf H.E. Boos,
V.V. Mangazeev,
S.M. Sergeev 
}\\
{\small Institute for High Energy Physics, 142284 Protvino, Moscow Region,
Russia}
\end{center}
\vspace{1cm}
\begin{center}
{\bf Abstract}
\end{center}
Using a modified version of the tetrahedron equations we construct a new
family
of $N$-state three-dimensional integrable models with commuting
two-layer transfer-matrices.
We investigate a particular class of solutions to these equations
and parameterize them in terms of elliptic functions. The corresponding
models contain one free parameter $k$ -- an elliptic modulus.

\newpage

\section{Introduction}

Recent progress in the construction of solvable lattice models
in higher dimensions has stimulated efforts to find new
interesting examples ``living'' in $D>2$ dimensions.
As is known the main ingredient of practically all two-dimensional
solvable models is a Yang-Baxter equation (YBE) which provides an existence
of a family of commuting transfer-matrices.
As a rule, the presence of such a family permits us to calculate the partition
function per site in the thermodynamical limit.
Unlike  two-dimensional conformal field theories the theory of the YBE
can be easily generalized for the cases $D>2$ \cite{Z1,Z2,B,BS,JM} and
it will lead us to the tetrahedron equations ($D=3$) and $D$-simplex equations.
The main problem, of course, is to solve them, because the number
of unknowns and equations grows with increasing dimension
in a catastrophic way.

Recently Bazhanov and Baxter in their paper \cite{BB} have generalized the
trigonometric Zamolodchikov model \cite{Z1,Z2,B} for the case of $N$ colours.
The weight functions of their model satisfy tetrahedron equations
(\cite{StSq}) and possess some remarkable symmetry properties.
But the essential shortcoming of Bazhanov-Baxter $N$-color model is an absence
any temperature-like parameter.

Authors of \cite{Man,NewTw} have proposed a new class of three-dimensional
models containing a temperature-like parameter $k$. In the limit $k\to 0$
the weight functions of these models reduce to the ``static limit'' case
of the Bazhanov-Baxter model.
The main idea of papers \cite{Man,NewTw} is to introduce a couple of modified
tetrahedron equations (MTE). In two dimensions this idea have been
considered in \cite{KS}. Then using a couple of modified
tetrahedron equations, the authors of \cite{Man,NewTw} have constructed
two-layer commuting transfer-matrices for the checkerboard lattice of
weights.

This paper, in fact, is a continuation of \cite{Man,NewTw}.
Our main purpose is to remove the ``static limit'' condition which is a
constraint on three ``angle'' parameters for weight functions.
The result is that each elementary
weight function of the model will depend on three spectral parameters
and one ``modulus''. Unfortunately, to  remove the ``static limit''
condition we were forced to
 refuse from the second modified tetrahedron equation.
It leads to a more complicated  commuting family
of two-lawer transfer-matrices.

The paper is organized as follows:
In Section 2 we give necessary definitions, the explicit expression
for weight functions and write out the modified tetrahedron equations.
In Section 3 we recall the proof of the modified tetrahedron equations
based on the Star-Square relation, give all necessary algebraic constraints
on the parameters and consider an important submanifold of ``equal''
moduli in the space of parameters (in next sections we restrict ourselves
to this case only). In Section 4 we introduce a parameterization of
weight functions in terms of angle-like variables. In fact, all the formulae
of Section 4 can be rewritten in terms of elliptic functions and
we give the corresponding formulae in the Appendix. Section 5 contains a
discussion of the symmetry properties of the weight functions under the group
of transformations of the elementary cube. In Section 6 we construct
a commuting family of two-layer transfer-matrices using the
solutions obtained from the modified tetrahedron equations and discuss the
structure of the lattice model. Finally, the Appendix contains
all the elliptic formulae and a detailed consideration of the $N=2$ case.

\section{Modified tetrahedron equations and
Body-Centered-Cube (BCC) ansatz for weight functions}

In this section we present the modified tetrahedron equations
(MTE) and briefly discuss their properties.
We recall necessary definitions from \cite{KMS,StSq,NewTw} and
give the symmetrical form of the weight functions $W$

Further we will follow a statistical mechanics interpretation and
use the notations introduced by Baxter in \cite{B} for the Zamolodchikov model.
Namely, consider a simple cubic lattice ${\cal L}$ and at each site of
${\cal L}$ place a spin variable taking its values in $Z_N$,
for any integer $N\ge2$.

To each elementary cube of the lattice we can assign some weight
function $W(a|efg|bcd|h)$, depending on eight surrounding spin
variables (see Fig.~1).

\begin{picture}(600,265)
\put(0,50){
\begin{picture}(500,200)
\multiput(140,0)(120,0){2}{\line(0,1){120}}
\multiput(140,0)(0,120){2}{\line(1,0){120}}
\multiput(140,0)(0,120){2}{\line(-1,1){60}}
\put(80,180){\line(1,0){120}}\put(80,180){\line(0,-1){120}}
\put(200,180){\line(1,-1){60}}
\multiput(200,180)(0,-20){6}{\line(0,-1){12}}
\multiput(80,60)(20,0){6}{\line(1,0){12}}
\multiput(255,5)(-30,30){2}{\line(-1,1){20}}
\multiput(140,0)(120,0){2}{\circle*{15}}
\multiput(140,120)(120,0){2}{\circle*{15}}
\multiput(80,60)(120,0){2}{\circle*{15}}
\multiput(80,180)(120,0){2}{\circle*{15}}
\put(300,80){\large $=\quad W(a|e,f,g|b,c,d|h)$}
\put(150,10){$e$}\put(270,10){$d$}\put(150,100){$a$}
\put(270,100){$f$}\put(212,186){$b$}\put(92,190){$g$}
\put(92,65){$c$}\put(212,65){$h$}
\end{picture}
}
\put(320,0){\Large\bf Fig. 1}
\end{picture}

Note that for different elementary cubes
of the lattice ${\cal L}$ we can use different weight functions
$W^{(1)}$, $W^{(2)}$ etc. Later we will fix the explicit structure
of the lattice ${\cal L}$, demanding that the two-layer transfer-matrices
commute between themselves.

Now let us consider two sets of weight functions: $W$, $W'$, $W''$, $W'''$
and $\ov W$, $\ov W'$, $\ov W''$, $\ov W'''$. Suppose that they satisfy
the following equations:
\begin{eqnarray}
\sum_{d}
&W(a_4|c_2,c_1,c_3|b_1,b_3,b_2|d)\ov W'(c_1|b_2,a_3,b_1|c_4,d,c_6|b_4)&
\nonumber\\
\times&W''(b_1|d,c_4,c_3|a_2,b_3,b_4|c_5)
\ov W'''(d|b_2,b_4,b_3|c_5,c_2,c_6|a_1)&\nonumber\\
=\sum_{d}
&W'''(b_1|c_1,c_4,c_3|a_2,a_4,a_3|d)\ov W''(c_1|b_2,a_3,a_4|d,c_2,c_6|a_1)&
\nonumber\\
\times&W'(a_4|c_2,d,c_3|a_2,b_3,a_1|c_5)\ov W(d|a_1,a_3,a_2|c_4,c_5,c_6|b_4),&
                                                             \label{2}
\end{eqnarray}
where $a_i,b_i,c_j\in Z_N$,
$i=1,\ldots,4$, $j=1,\ldots,6$. We will call relations (\ref{2}) the
modified tetrahedron equations. Note that all eight weights in (\ref{2})
are independent, but we placed  $W$ and $\ov W$ in relations (\ref{2})
so as to follow the notation of papers \cite{Man,NewTw}.
For the case $W=\ov W$ we come to the
usual tetrahedorn equations (see, for example, \cite{B}).
We would like to stress that contrary to \cite{Man,NewTw} we do not demand
the validity of the dual variant of (\ref{2}) (with all $W$'s replaced
by $\ov W$'s and vice versa).
The absence of the dual variant of (\ref{2}) permits us to overcome the static
limit condition (see \cite{Man,NewTw}) but leads to a more complicated
commutativity of the two-layer transfer-matrices.

Let us recall some definitions used in papers \cite{KMS,StSq,NewTw}.
First denote
\begin{equation}
\w=\exp(2\pi i/N),\qquad \w^{1/2}=\exp(\pi i/N).                \label{4}
\end{equation}
Further, taking $x,y,z$ to be complex parameters constrained by the
Fermat equation
\begin{equation}
x^N+y^N=z^N                                                    \label{5}
\end{equation}
and $l$ to be an element of $Z_N$, define
\begin{equation}
w(x,y,z|l)=\prod_{s=1}^{l}{y\over z-x\w^s}.                    \label{6}
\end{equation}
In addition, define the function with one more argument
\begin{equation}
w(x,y,z|k,l)=w(x,y,z|k-l)\Phi(l),\quad k,l\in Z_N,               \label{7}
\end{equation}
where
\begin{equation}
\Phi(l)=\w^{l(l+N)/2}.                                          \label{8}
\end{equation}
Let us mention also two formulae for the $w$ functions, which are useful for
calculations:
\begin{equation}
w(x,y,z|l+k)=w(x,y,z|k)w(x\w^k,y,z|l),                         \label{9}
\end{equation}
\begin{equation}
w(x,y,z|k,l)=\w^{kl}/w(z,\w^{1/2}y,\w x|l,k).                  \label{10}
\end{equation}

Now introduce the set of homogeneous variables $x_i$, $i=1,\ldots,8$ and
$x_{13}$, $x_{24}$, $x_{58}$, $x_{67}$ satisfying
\begin{equation}
x_{13}^N=x_1^N-x_3^N,\quad x_{24}^N=x_2^N-x_4^N,\quad
x_{58}^N=x_5^N-x_8^N,\quad x_{67}^N=x_6^N-x_7^N.                \label{11}
\end{equation}
We will also need six additional variables to ensure the necessary
transformation properties of the weight functions under the group $G$
of transformations of a three-dimensional cube. Define them as
\begin{eqnarray}
&u^N=(x_3x_5/x_1)^N-x_8^N,\quad v^N=x_7^N-(x_4x_6/x_2)^N,&\nonumber\\
&\xi^N=(x_1x_7/x_3)^N-x_6^N,\quad\l^N= x_5^N-(x_2x_8/x_4)^N,&\nonumber\\
&\mu^N=(x_{13}x_{24}x_7/x_3x_4)^N-(x_{58}x_{67}/x_8)^N,&\nonumber\\
&\nu^N=(x_{13}x_{24}x_6/x_1x_2)^N-(x_{58}x_{67}/x_5)^N.&\label{211}
\end{eqnarray}

Using these definitions we define the weight function $W(a|efg|bcd|h)$
as
\begin{eqnarray}
&&W(a|e,f,g|b,c,d|h)=
\biggl[\frac {\di w(x_{58}x_{67},x_8\mu,\,
x_{13}x_{24}x_7x_8/x_3x_4|a+d,e+f)} {
\di w(x_{58}x_{67},x_5\nu,\,x_{13}x_{24}x_5x_6/x_1x_2|g+h,b+c)}
\biggr]^{1/2}\nonumber\\
&&\times\biggl[{\di w(x_1x_8,x_1u,x_3x_5|e+h,d+c)\over
\di w(x_4x_6,x_2v,x_2x_7|a+b,f+g)}\biggr]^{1/2}
\biggl[{\di w(x_2x_8,x_4\l,x_4x_5|e+g,a+c)\over
\di w(x_3x_6,x_3\xi,x_1x_7|b+d,f+h)}\biggr]^{1/2}\nonumber\\
&&\times
{\di \w^{bf}\w^{(ag+gb+bh)/2}\over\di \w^{ag}\w^{(dh+de+ea)/2}}
\biggl\{{\di\sum_{\sigma\in Z_N}}{\di w(x_3,x_{13},x_1|d,h+\sigma)
w(x_4,x_{24},x_2|a,g+\sigma)\over
\di w(x_8,x_{58},x_5|e,c+\sigma)w(x_7/\w,x_{67},x_6|f,b+\sigma)}\biggr\}_0
\nonumber\\                                              \label{12}
\end{eqnarray}
where the lower index ``$0$'' after the curly brackets implies
that the expression in the curly brackets is divided by itself with all
exterior spin variables equated to zero.

Note that all gauge multipliers (face and edge types) before curly brackets
in (\ref{12}) are necessary in order to provide the correct symmetry properties
of the weight functions under the transformations of the group $G$.

Formula (\ref{12}) generalizes the weight functions of the model
proposed by Bazhanov and Baxter \cite{BB,KMS}. In fact, it coincides
with the weight functions from \cite{NewTw} up to gauge multipliers and
we will also call (\ref{12}) the Body-Centered-Cube (BCC) ansatz for
weight functions. But in contrast to \cite{NewTw} we will introduce
a new parameterization for $x_i$ in a such way that each weight function
will depend on three independent spectral parameters and modulus
parameter $k$.

\section{The proof of the modified tetrahedron equations}

In this section we recall the proof of the modified tetrahedron
equations
(\ref{2}) for weight functions (\ref{12}). In fact, it reproduces
the method of papers \cite{StSq,NewTw} and we give it here for completeness.

The main idea of papers \cite{StSq,NewTw} is to reduce complicated
modified tetrahedron equations (\ref{2}) to a couple of much more
simple relations: namely ``inversion'' and Star-Square ones.

The ``inversion'' relation for functions $w(x,y,z|k,l)$ has the form:
\begin{equation}
\sum_{k\in Z_N}{w(x,y,z|k,l)\over w(x,y,\w z|k,m)}=
N\delta_{l,m}{(1-z/x)\over (1-z^N/x^N)},                    \label{13}
\end{equation}
where $l,\>m\in Z_N$, $x,y,z$ satisfy (\ref{5}) and $\delta_{l,m}$ is
the Kronecker symbol on $Z_N$.

The Star-Square relation permits us to calculate the sum over the one spin
variable from the product of four $w$ functions provided that their
arguments satisfy some algebraic constraint.
To avoid the introduction of additional variables for ensuring the
cyclic property of all $w$'s {\it modulo} $N$, we will use
a non-cyclic analog of $w$ function, defined recurrently as follows:
\begin{equation}
w(x|0)=1,\quad{w(x|l)\over w(x|l-1)}={1\over(1-x\w^l)},\quad l\in Z,
                                                            \label{14}
\end{equation}
where $Z$ is the set of all integers.
It is obvious that
\begin{equation}
w(x,y,z|l)=
{\Bigl(}{y\over z}{\Bigr)}^l w(x/z|l),\quad l\in Z_N, \label{15}
\end{equation}
where index $l$, being considered modulo $N$, is interpreted as an element
of $Z_N$.

Then the Star-Square relation can be written as:
\begin{eqnarray}
&\biggl\{{\di
\sum_{\s\in Z_N} {{w(x_1,y_1,z_1|a+\s)w(x_2,y_2,z_2|b+\s)}\over
{w(x_3,y_3,z_3|c+\s)w(x_4,y_4,z_4|d+\s)}}}\biggr\}_0=&\nonumber\\
&{\di
{(x_2z_1/x_1z_2)^{a-b}(x_1y_2/x_2z_1)^b(z_3/y_3)^c(z_4/y_4)^d\over
\Phi (a-b)\w^{(a+b)/2}}}&\nonumber\\
&\times{\di{w(\w x_3x_4z_1z_2/x_1x_2z_3z_4|c+d-a-b)}\over
\di{w({x_4z_1\over x_1z_4}|d-a)w({x_3z_2\over x_2z_3}|c-b)
w({x_3z_1\over x_1z_3}|c-a)w({x_4z_2\over x_2z_4}|d-b)}}&,
                                                               \label{16}
\end{eqnarray}
where the lower index ``$0$'' after the curly brackets indicates that the
l.h.s. of (\ref{16}) is normalized to unity at zero exterior spins, and
the following constraint is imposed
\begin{equation}
{y_1\over z_1}{y_2\over z_2}{z_3\over y_3}{z_4\over y_4}=\w.    \label{17}
\end{equation}
Note that the separate $w$'s in the r.h.s. of (\ref{16})
are not single-valued functions on $Z_N$, while the whole
expression is cyclic in the exterior spins $a,b,c,d$.

Relations (\ref{13}), (\ref{16}) are proved by using  properties
of Fourier transformation over spin variables and the detailed proof was
given in \cite{StSq}.

Now we will prove relations (\ref{2}) for ansatz (\ref{12})
with some appropriate choice of constant multiplier $R$.
Instead of weights $W$, $W'$, $W''$, $W'''$ let us substitute into (\ref{2})
explicit formula (\ref{12}) with corresponding sets of parameters:
\begin{eqnarray}
W,W',W'',W'''\to W(x_i,x_{ij}), W(x'_i,x'_{ij}),
W(x''_i,x''_{ij}), W(x'''_i,x'''_{ij}),&&\nonumber\\
\ov W,\ov W',\ov W'',\ov W'''\to W(\ov x_i,\ov x_{ij}),W(\ov x'_i,\ov x'_{ij}),
W(\ov x''_i,\ov x''_{ij}),W(\ov x'''_i,\ov x'''_{ij}).&&\label{18}
\end{eqnarray}
Now we will show that all equations (\ref{2}) are equivalent to some
algebraic system of nonlinear equations for parameters $x_i$, $\ov x_i$
etc.

Hereafter we will fix a normalization of all parameters $x's$ that enter
in the definition (\ref{12}) as
\begin{equation}
x_3=1,\quad x_4=1,\quad x_7=1,\quad x_8=1,           \label{19}
\end{equation}
and similarly for all the $\ov x's$.

It is easy to see that after substituting of (\ref{12})
in relations (\ref{2}) we will explicitly obtain 24 pairs of $w$ functions
coming from the multipliers before the curly brackets in (\ref{12}).
Let us demand that all these factors cancel each other (for each $w$ function
there is one partner with the same dependence from spin variables).
Then we obtain 24 relations for variables $x_i$, $x_{ij}$ and $\ov x's$,
$\ov x_{ij}'s$ (only 21 of the constraints are independent)
and 24 relations for the additional variables defined by (\ref{211}).
In fact, they relate the set of rapidity like parameters between
different weight functions.

Let us multiply both sides of (\ref{2}) by the following product of $w$
weights:
\begin{eqnarray}
&{\di w(x''_7,x''_{67},x''_6|c_4,a_2+l_2)\over\di
w(\ov x'''_3/\w,\ov x'''_{13},
\ov x'''_1|c_6,a_1+l_1)}{\di
w(\ov x'_7,\w\ov x'_{67},\ov x'_6|a_3,c_4+l_3)\over
\di w(x_4,x_{24},\w x_2|a_4,c_3+l_4)}&\times\nonumber\\
\times&{\di w(\ov x''_8,\ov x''_{58},\ov x''_5/\w|b_2,c_2+m_2)\over
\di w(x'''_4,
x'''_{24},\w x'''_2|b_1,c_3+m_1)}{\di w(\w x'_8,x'_{58},x'_5|c_2,b_3+m_3)\over
\di w(\ov x_3/\w,\ov x_{13},\ov x_1|c_6,b_4+m_4)}&                  \label{19a}
\end{eqnarray}
and sum over $a_1,a_2,a_3,a_4,\>b_1,b_2,b_3,b_4$.
Note that due to  the ``inversion'' relation (\ref{13}) we do not
lose any information after such transition of spins
$a_i,b_i$ to $l_i,m_i$.

The functions $w$ in expression (\ref{19}) are chosen in such a way
that using relation (\ref{13}) we can calculate the sums over the spins
$a_1,a_2,a_3,a_4$ in the l.h.s. and those over $b_1,b_2,b_3,b_4$ in the
r.h.s. of the obtained equation and cancel the summations over the
spins $\sigma$'s, which come from expression (\ref{12})
for the functions $W$'s and $\ov W$'s.
Now let us consider the applicability conditions (\ref{17}) of
Star-Square relation (\ref{16}) for the sums over $a_1,a_2,a_3,a_4$
in the r.h.s. and over $b_1,b_2,b_3,b_4$ in the l.h.s. of the obtained
equation. We obtain eight conditions on the $x$'s and $\ov x$'s.
Applying relation (\ref{16}) eight times and calculating the sums
over $d$ spin in the l.h.s. and r.h.s (the spin structure of the sums
over $d$ has the form of relation (\ref{13}) and we demand that
corresponding variables $x,y,z$ entering in the arguments of functions
$w(x,y,z|k,l)$ are constrained in  such a way that relation
(\ref{13}) can be used), we come to the equation without any summation.
The l.h.s. and r.h.s. of this equation consists of the products of ten $w$
functions  with the same spin structure and
expressions like $x_1^{m_2}$ coming from relation (\ref{16}).
Let us impose all the necessary constraints on parameters $x_i$, $x_{ij}$,
$\ov x_i$ and $\ov x_{ij}$ to satisfy this equation.
The explicit calculations show that we obtain 11 more independent
constraints on parameters. Taking into account 21+24=45 constraints
coming from a cancellation of ``face'' factors in (\ref{12}) we
have an algebraic system of 56 nonlinear equations.

Taking into account relations (\ref{11},\ref{211}) it is easy to understand
that only 23 of the constraints are independent
on the level of $N$-th powers of coordinates $x's$ and $\ov x's$.
Since each weight of form (\ref{12}) depends only on four independent
parameters, we have a nine parameter solution of relations (\ref{2}).

To order all these constraints let us introduce the following rapidity
like parameters:
\begin{equation}
a={\di x_5x_6\over x_1x_2},\quad
i_1=\w{\di x_6\over\di x_2},\quad i_2={\di x_5\over x_2},\quad
i_3=a{\di x_{13}x_{24}\over\di x_{58}x_{67}},                \label{20}
\end{equation}
\begin{equation}
j_1=\w{\di x_1\over\di x_5},\quad j_2={\di x_1\over x_6},\quad
j_3={\di x_{13}x_{24}\over x_{58}x_{67}}.                     \label{21}
\end{equation}
and the same for the $\ov x's$.
Then the set of 24 relations following from a cancellation of ``face''
factors (which do not contain additional variables defined in (\ref{211}))
has the form:
\begin{equation}
i_1=i_2''',\quad i_2=i_2',\quad i_3=i_2'',\quad i_1'=i_3''',\quad
i_3'=i_3'',\quad i_1''=i_1''',                              \label{24}
\end{equation}
\begin{equation}
\ov j_1=\ov j_2''',\quad \ov j_2=\ov j_2',\quad \ov j_3=\ov j_2'',
\quad \ov j_1'=\ov j_3''',\quad\ov j_3'=\ov j_3'',
\quad \ov j_1''=\ov j_1''',                              \label{25}
\end{equation}
\begin{eqnarray}
j_1=\ov i_2''',\quad j_2=\ov i_2',\quad j_3=\ov i_2'',\quad j_1'=\ov i_3''',
\quad j_3'=\ov i_3'',\quad j_1''=\ov i_1''',                \nonumber\\
\ov i_1=j_2''',\quad \ov i_2=j_2',\quad \ov i_3=j_2'',\quad \ov i_1'=j_3''',
\quad \ov i_3'=j_3'',\quad \ov i_1''=j_1'''.                \label{26}
\end{eqnarray}
Only 21 of the 24 relations (\ref{24}-\ref{26}) are
independent. A simple consequence of (\ref{24}-\ref{25}) is
\begin{equation}
\Delta\equiv{\di a\over \di\ov a}={\di a'\over \di\ov a'}=
{\di a''\over \di\ov a''}={\di a'''\over \di\ov a'''}       \label{27}
\end{equation}
and we can conclude that a parameter $\Delta$ is an absolute invariant of
relations (\ref{2}).

We will also define rapidity variables
of the mixed type:
\begin{equation}
\a_1={\di x_{13}\ov x_{67}\over \di x_{67}\ov x_{13}},\quad
\a_2={\di x_{13}\ov x_{58}\over \di x_{58}\ov x_{13}},\quad
\a_3={\di x_1\over\di \ov x_1}.                               \label{28}
\end{equation}
Then six of the eleven additional relations coming from the coincidence
of the structures of l.h.s. and r.h.s. of the modified tetrahedron equations
(after using of ``inversion'' and Star-Square relations)
take the form
\begin{equation}
\a_1=\a_2''',\quad \a_2=\a_2',\quad \a_3=\a_2'',\quad \a_1'=\a_3''',
\quad \a_3'=\a_3'',\quad \a_1''=\a_1'''.                 \label{29}
\end{equation}
Now we can similarly rewrite 24 of the constraints containing additional
variables
defined in (\ref{211}). Define
\begin{equation}
\b_1=\w^{1/2}v,\quad \b_2={\di \l\over\di x_2},\quad \b_3={\di\nu x_5\over
\di x_{58}x_{67}},                                       \label{30}
\end{equation}
\begin{equation}
\g_1=\w^{1/2}{\di ux_1\over\di x_5},\quad \g_2={\di\xi\over\di x_6},\quad
\g_3={\di\mu\over\di x_{58}x_{67}}                      \label{31}
\end{equation}
and similarly for $\ov \b_i$, $\ov \g_i$. Then we have
\begin{equation}
\b_1=\b_2''',\quad \b_2=\b_2',\quad \b_3=\b_2'',\quad \b_1'=\b_3''',\quad
\b_3'=\b_3'',\quad \b_1''=\b_1''',                              \label{32}
\end{equation}
\begin{equation}
\ov \g_1=\ov \g_2''',\quad \ov \g_2=\ov \g_2',\quad \ov \g_3=\ov \g_2'',
\quad \ov \g_1'=\ov \g_3''',\quad\ov \g_3'=\ov \g_3'',
\quad \ov \g_1''=\ov \g_1'''                              \label{33}
\end{equation}
and
\begin{eqnarray}
\g_1=\ov \b_2''',\quad \g_2=\ov \b_2',\quad \g_3=\ov \b_2'',\quad
\g_1'=\ov \b_3''',\quad \g_3'=\ov \b_3'',\quad \g_1''=\ov \b_1''', \nonumber\\
\ov \b_1=\g_2''',\quad \ov \b_2=\g_2',\quad \ov \b_3=\g_2'',\quad
\ov \b_1'=\g_3''',\quad \ov \b_3'=\g_3'',\quad \ov \b_1''=\g_1'''.  \label{34}
\end{eqnarray}
The remaining five constraints include parameters of all four sets
of weight functions and correspond to a generalization of tetrahedron
quadrilateral constraint for the Zamolodchikov model.
We can choose them as
\begin{equation}
x_{24}x_{24}''=x_{24}'x_{24}'''                              \label{35}
\end{equation}
and
\begin{equation}
{\di \ov x_{13}\over\di\ov x_1}{\di\ov x_1'\over\di\ov x_{13}'}
{\di \ov x_{13}''\over\di\ov x_1''}{\di\ov x_1'''\over\di\ov x_{13}'''}=1,
\quad \w{\di x_{67}\over\di\ov x_{24}'}{\di x_{24}''\over\di x_2''}
{\di x_2'''\over\di x_{24}'''}=1,\quad
\w {\di x_{67}'\over\di\ov x_{24}}{\di x_{67}'''\over\di x_{67}''}=1,\quad
{\di x_{58}\over\di x_{58}'}{\di x_{58}''\over\di\ov x_{24}'''}=1. \label{36}
\end{equation}

Now let us briefly discuss the main properties of relations
(\ref{24}-\ref{36}). As we mention all these equations have a nine
parametric solution. We can construct it in the following way.
Relations (\ref{24},\ref{35}) give 7 restrictions on 16 parameters
$x_i$, $x_i'$, $x_i''$ and $x_i'''$. Then all parameters $\ov x_i$ can be
calculated in a unique way using relations
(\ref{25},\ref{26},\ref{29},\ref{35}). All remaining constraints will be
satisfied automatically up to some $N$-th root of unity. And choosing
the degrees of the $\w$'s in a corresponding way we will satisfy all relations
(\ref{24}-\ref{36}) and therefore the modified tetrahedron equation (\ref{2}).

Now let us consider an important particular case of relations
(\ref{24}-\ref{36}):
\begin{equation}
a=a'=a''=a'''.                                  \label{38}
\end{equation}
Constraints (\ref{38}) reduce the number of parameters to six.
In this paper we will consider only a class of solutions for
the modified tetrahedron equations restricted
by conditions (\ref{38}).
Then the parameter $a$ can be interpreted as an invariant.
Detailed analysis of equations (\ref{24},\ref{35})
shows that on the surface (\ref{38}) we have two solutions.
Only the first one contains the Zamolodchikov -- Bazhanov -- Baxter
model in the limit $a\to1$ .
So we will investigate this case only. Then one can show that
\begin{equation}
\Delta=a^2={\di 1\over\di\ov a^2}.               \label{39}
\end{equation}
It is easy also to show that the following relations between the $N$-th powers
of the parameters $i_k$, $j_k$, $\a_k$, $\b_k$ and $\g_k$ are valid:
\begin{equation}
j_k^N={\di i_k^N\over a^N},\quad \a_k^N={\di j_k^N-1\over\di i_k^N-1},\quad
\b_k^N=i_k^N-1,\quad \g_k^N=j_k^N-1,\quad k=1,2,3.   \label{40}
\end{equation}
Note that formulae (\ref{40}) for the $N$-th powers of parameters $\a_k$
give a convenient way to find all parameters $\ov x's$ in terms of $x's$.

In the next section we will give a convenient parameterization
of all parameters in terms of variables resembling tetrahedron angles.

\section{Parameterization}

Define $m,T_1,T_2,T_3$ through the relations
\begin{eqnarray}
&{\di x_2^N\over\di x_6^N} = -mT_1^2;\quad
{\di x_2^N\over\di x_5^N} = -mT_2^2;&\nonumber\\
&{\di x_{13}^Nx_{24}^N\over\di x_{58}^N x_{67}^N} = -mT_3^2;\quad
{\di x_1^N x_2^N\over\di x_5^N x_6^N} = m^2.&\label{par10}
\end{eqnarray}
The advantage of this parameterization is that when we put $m=1$
and $T_i=\tan(\theta_i/2)$, we obtain the conventional
parameterization of the
Zamolodchikov -- Bazhanov -- Baxter model.

Solving (\ref{par10}) with respect to the $x_i^N$'s, we obtain
\begin{eqnarray}
&x_1^N = {\di 1\over\di T_1T_2}\di\sqrt{\di 1+mT_3^2\over\di 1+m^{-1}T_3^2}
\exp (ia_3),&\nonumber\\
&x_2^N = T_1T_2\di\sqrt{\di 1+mT_3^2\over\di 1+m^{-1}T_3^2}
\exp (ia_3),&\nonumber\\
&x_5^N = -m^{-1}{\di T_1\over\di T_2}\di\sqrt{\di 1+mT_3^2\over\di
1+m^{-1}T_3^2}
\exp (ia_3),&\nonumber\\
&x_6^N = -m^{-1}{\di T_2\over\di T_1}\di\sqrt{\di 1+mT_3^2\over\di
1+m^{-1}T_3^2}
\exp (ia_3),&\label{par11}
\end{eqnarray}
where $a_3$ is defined from its cosine:
\begin{equation}
\cos (a_3) = {\di 1+T_1^2T_2^2-T_1^2T_3^2-T_2^2T_3^2\over
\di 2 T_1T_2\sqrt{\di (1+mT_3^2)(1+m^{-1}T_3^2)}}.\label{par12}
\end{equation}
The formulae for the $\ov x_i$'s can be obtained from these ones by changing
$m\rightarrow m^{-1}$, $a_3$ being the same.

The formula (\ref{par12}) can be rewritten in another form by the
substitution
\begin{equation}
T_{p}^2 = {D_{p}-C_{p}\over D_{p}+C_{p}} =
({k'S_{p}\over D_{p}+C_{p}})^2\label{par13}
\end{equation}
where
\begin{eqnarray}
&D_{p}^2 = 1-k^2S_{p}^2,\quad C_{p}^2 = 1 - S_{p}^2,&\nonumber\\
&m = {\di 1-k\over\di 1+k},\quad k^2+k'^2 = 1.&\label{par14}
\end{eqnarray}
and $p = 1,2,3$.
Then we have (defining the angles $a_1$ and $a_2$ at once)
\begin{equation}
\cos (a_r) = {\di D_pD_qC_r+C_pC_qD_r\over \di k'^2 S_pS_q},\label{par15}
\end{equation}
where $\{p,q,r\} = \{1,2,3\}$. Introduce else the angle $a_0$ as follows:
\begin{equation}
\cos (a_0) = {\di D_1D_2D_3+k^2C_1C_2C_3\over \di k'^2 }.\label{par16}
\end{equation}
$a_0$ is symmetric under the permutation of $T_1,T_2,T_3$, and the sign of
$a_0$ we choose as follows
\begin{equation}
\sin (a_0) = k \sin(a_r) S_p S_q.\label{par17}
\end{equation}
To write down $x_{ij}^N$ we need to introduce a ``linear excesses''
\begin{eqnarray}
&\di\beta_0 = \pi-{\di a_1+a_2+a_3+a_0\over\di 2},\quad
\beta_r = {\di a_p+a_q-a_r-a_0\over\di 2},&\nonumber\\
&\di\ov \beta_0 = \pi-{\di a_1+a_2+a_3-a_0\over\di 2},\quad
\ov \beta_r = {\di a_p+a_q-a_r+a_0\over\di 2}.&\label{par18}
\end{eqnarray}
Then
\begin{eqnarray}
&x_{13}^N = - m^{1/2}{\di T_3\over\di T_1T_2}
\di\sqrt{\di (1+m^{-1}T_1^2)(1+m^{-1}T_2^2)\over\di (1+m^{-1}T_3^2)}
\exp (-i\beta_0),&\nonumber\\
&x_{24}^N = - m^{-1/2} T_3
\di\sqrt{\di (1+mT_1^2)(1+mT_2^2)\over\di (1+m^{-1}T_3^2)}
\exp (-i\beta_3),&\nonumber\\
&x_{58}^N = - m^{-1/2}{\di 1\over\di T_2}
\di\sqrt{\di (1+m^{-1}T_1^2)(1+mT_2^2)\over\di (1+m^{-1}T_3^2)}
\exp (i\ov\beta_2),&\nonumber\\
&x_{67}^N = - m^{-1/2}{\di 1\over\di T_1}
\di\sqrt{\di (1+mT_1^2)(1+m^{-1}T_2^2)\over\di (1+m^{-1}T_3^2)}
\exp (i\ov\beta_1),&\label{par19}
\end{eqnarray}
We mention again that the $\ov x_i^N$'s and $\ov x_{ij}^N$'s
can be obtained by the same formulae as
$x_i^N$ and $x_{ij}^N$ with $m$ replaced by $m^{-1}$
(hence $k$ replaced by $-k$ and $a_0$ replaced by $-a_0$),
$a_{1,2,3}$ being unchanged.

One can take the $N$-th roots of the expressions (\ref{par11},\ref{par19}) so
that when $m\rightarrow 1$ there appears the parameterization of the
Zamolodchikov -- Bazhanov -- Baxter
model. Namely, if $m$, $T_r$ and $a_r$, $r=1,2,3$, are real positive then
we can choose the phases as follows
\begin{eqnarray}
&x_1=|x_1|\exp(ia_3/N),\quad x_{13}=|x_{13}|\w^{1/2}\exp(-i\beta_0/N);
&\nonumber\\
&x_2=|x_2|\exp(ia_3/N),\quad x_{24}=|x_{24}|\w^{1/2}\exp(-i\beta_3/N);
&\nonumber\\
&x_5=|x_5|\w^{1/2}\exp(ia_3/N),\quad
x_{58}=|x_{58}|\w^{1/2}\exp(i\ov\beta_2/N);
&\nonumber\\
&x_6=|x_6|\w^{-1/2}\exp(ia_3/N),\quad
x_{67}=|x_{67}|\w^{-1/2}\exp(i\ov\beta_1/N).
&\label{par20}
\end{eqnarray}
In addition we reproduce some exotic formulae for the ``spherical sides''.
\begin{eqnarray}
&\di\sqrt{\di\sin\beta_0\sin\beta_1\sin\beta_2\sin\beta_3\over
\di\sin\ov\beta_0\sin\ov\beta_1\sin\ov\beta_2\sin\ov\beta_3}=m,&
\nonumber\\
&\di mT_r^2 = {\di\sin\beta_p\sin\beta_q\over\di
\sin\ov\beta_0\sin\ov\beta_r},&\nonumber\\
&\di kS_r^2 = {\di\sin a_0\sin a_r\over\di\sin a_p\sin a_q}.&\label{par21}
\end{eqnarray}

Now rewrite the scheme of arguments of weights $(W,\ov W)$
in the modified tetrahedron equations.
Exhibiting the dependence on $T_{1,2,3}$ and corresponding
$a_{0,1,2,3}$ as
\begin{equation}
W =W(T_1,T_2,T_3)\rightarrow a_0,a_1,a_2,a_3,\label{par22}
\end{equation}
the arguments of the weights in the modified tetrahedron equations
can be written as follows
\begin{eqnarray}
&W = W(T_1,T_2,T_3)\rightarrow a_0,a_1,a_2,a_3;&\nonumber\\
&W' = W(T_4,T_2,T_5)\rightarrow a'_0,a'_1,a'_2,a'_3;&\nonumber\\
&W'' = W(T_6,1/T_3,T_5)\rightarrow a''_0,a''_1,a''_2,a''_3;&\nonumber\\
&W''' = W(T_6,T_1,1/T_4)\rightarrow a'''_0,a'''_1,a'''_2,a'''_3.&\label{par23}
\end{eqnarray}
In terms of the ``spherical sides'' of the ``tetrahedron'', the modified
tetrahedron equations  correspond to the set of equivalent
relations
\begin{eqnarray}
&a_0+a'_2+a'''_2 =a''_2  ,\quad a'_0+a_2+a'''_3 = a''_3  ,&\nonumber\\
&a''_0+a_3+a'''_1 = a'_3  ,\quad a'''_0+a_1 +a''_1 = a'_1  &\label{par24}
\end{eqnarray}
Concluding this section we note that some of the formulae
resemble those for the elliptic functions. Surely one can put
$S = \sn(\theta,k)$ for $S$ and $k$ defined by (\ref{par13},\ref{par14}).
Such version of the formulae we shall collect in the Appendix, where we also
give the explicit form of the Boltzmann weight functions for the case $N=2$.

\section{Symmetry properties}

In this section we give all the necessary formulae for the transformation
of parameters from which the weight function (\ref{12}) depends on
under the action of elements of the group $G$ of all symmetry
transformations of the three-dimensional cube. The main advantage of
introducing parameters (\ref{11}-\ref{211}) is that any element
from group $G$ induces a multiplicative law of the transformation of
parameters $x_i$, $x_{ij}$, $u$, $v$, $\xi$, $\l$, $\mu$ and $\nu$.

Any element of the group $G$ can be represented as a composition
of two generating elements: $\tau$ and $\rho$ (see \cite{KMS,BB2}).
The action of these two elements on the set of spins $\{a|e,f,g|b,c,d|h\}$
(see Fig.~1) can be described as follows
\begin{equation}
\tau\{a|e,f,g|b,c,d|h\}=\{a|f,e,g|c,b,d|h\} \label{sym1}
\end{equation}
and
\begin{equation}
\rho\{a|e,f,g|b,c,d|h\}=\{g|c,a,b|f,h,e,|d\}. \label{sym2}
\end{equation}
In fact, $\tau$ corresponds to the reflection of a three-dimensional cube
in the plane $\{aghd\}$ and $\rho$ -- to a rotation of $\pi/2$
around the vertical axis of the cube.
Further it will be convenient to remove normalization conditions (\ref{21})
and to restore the homogeneity of the parameters $x$'s.

The action of the element $\rho$ can be obtained by use
of the Fourier transformation (see \cite{KMS}) and the action of
$\tau$ is a quite obvious. Here we give only the resulting formulae
for their action on the coordinates. Note that all coordinates can be
split into four multiplets: $(x_3,x_{13},x_1)$, $(x_4,x_{24},x_2)$,
$(x_8,x_{58},x_5,u,\l)$ and $(x_7,x_{67},x_6,v,\xi,\mu,\nu)$ and that
any group element acts homogeneously on these four sets.
Also, we must demand a trivial action of the elements: $\tau^2$, $\rho^4$ and
$(\tau\rho)^6$ (trivial transformations) on all coordinates.
This will fix some uncertainties in the $N$-th roots of unity
in the transformation laws.

Then a correct action of the elements $\tau$ and $\rho$ looks like:
\begin{eqnarray}
&(x_3,x_{13},x_1)\stackrel{\tau}\rightarrow(x_3,x_{13},x_1),\quad
(x_4,x_{24},x_2)\stackrel{\tau}\rightarrow(x_4,x_{24},x_2),&\nonumber\\
&(x_8,x_{58},x_5,u,\l)\stackrel{\tau}\rightarrow
(x_7/\w,x_{67},x_6,\w^{-1/2}{\di x_3\xi\over\di x_1},
\w^{-1/2}{\di x_2v\over\di x_4}),&\label{sym3}\\
&(x_7,x_{67},x_6,v,\xi,\mu,\nu)
\stackrel{\tau}\rightarrow(\w x_8,x_{58},x_5,\w^{1/2}{\di x_4\l\over\di x_2},
\w^{1/2}{\di x_1u\over\di x_3},\w{\di x_8\mu\over\di x_7},
{\di x_5\nu\over\di x_6})&  \nonumber
\end{eqnarray}
and
\begin{eqnarray}
&(x_3,x_{13},x_1)\stackrel{\rho}\rightarrow(x_8x_{13},x_1u,x_3x_{58}),\quad
(x_4,x_{24},x_2)\stackrel{\rho}\rightarrow(x_2x_{67},x_2v,x_6x_{24}),&
\nonumber\\
&(x_8,x_{58},x_5,u,\l)\stackrel{\rho}\rightarrow
(x_5x_{13},\w x_1u,\w x_1x_{58},\w^{1/2}{\di x_1x_{13}u\over\di x_3},
\w^{1/2}{\di x_1x_5\nu\over\di x_{67}}),&\label{sym4}\\
&(x_7,x_{67},x_6,v,\xi,\mu,\nu)
\stackrel{\rho}\rightarrow&\nonumber\\
&(\w x_4x_{67},x_2v,x_7x_{24},
\w^{1/2}{\di x_2x_{67}v\over\di x_6},\w^{1/2}{\di x_3x_4\mu\over\di x_{13}},
\w{\di x_1x_4uv\l\over\di x_5x_8x_{13}},
{\di x_2uv\xi\over\di x_6x_{58}}).&  \nonumber
\end{eqnarray}

Substituting formulae (\ref{sym1}-\ref{sym4}) in (\ref{12}) one can show
that the whole $W$ function will be invariant
under $\tau$ and $\rho$ transformations from the group $G$.
The gauge factors before curly brackets in (\ref{12}) are chosen in a such
way to cancel multipliers coming after the Fourier transformation
of the sum in (\ref{12}).
The action of any element from $G$ on coordinates can be easily calculated
by a composition of formulae (\ref{sym3},\ref{sym4}).

Let us briefly comment on these formulae. It is easy to see that after
the $\rho$ transformation, the modulus parameter $a$ (see \ref{20}) transforms
as: $a\stackrel{\rho}\rightarrow a^{-1}$. So it is not invariant under
elements from the group $G$ containing an odd number of $\rho$
transformations. In some sense this is a pseudo-invariant (it is obvious
that any combination like $a+ a^{-1}$ will be invariant under all
transformations from $G$). Introducing ``angles'' as  arguments
of the elliptic functions (see Appendix) we can define a ``crossing''
transformation which in fact corresponds to some spin permutation like
$\{a|efg|bcd|h\}\to\{a|fge|cdb|h\}$) and etc.

In the next section we will show that the weight functions of the lattice
should have alternating values of moduli: $a$ and $a^{-1}$ in all directions
in order to provide a commuting family of two-layer transfer-matrices.

\section{Two-layer model}

In this section we will construct a commuting family of
two-layer transfer-matrices from solutions of the modified tetrahedron
equations (\ref{2}).
More explicitly, we will show that a composite weight function consisting
of eight elementary $W$ satisfying (\ref{2}) can be chosen in a such way
that it will satisfy
the tetrahedron equations of mixed type with spin variables placed in
the corner sites, in the middles of edges and in the centers of all faces
of the composite cube.

Consider a composite weight ${\cal W}$ which consists of eight independent
weight functions $W_a$, $W_e$, $W_f$, $W_g$, $W_b$, $W_c$, $W_d$, $W_h$
(see Fig.~6.1) and the same for ${\cal W'}$, ${\cal W''}$ and ${\cal W'''}$.

\begin{picture}(600,400)
\put(200,65){
\begin{picture}(300,300)
\multiput(0,100)(0,110){3}{\line(5,-4){125}}
\multiput(220,320)(-110,0){2}{\line(5,-4){125}}
\multiput(0,100)(62.5,-50){2}{\line(0,1){220}}
\multiput(0,320)(62.5,-50){2}{\line(1,0){220}}
\multiput(125,0)(0,110){3}{\line(1,0){220}}
\multiput(125,0)(110,0){3}{\line(0,1){220}}
\multiput(5,100)(20,0){11}{\line(1,0){10}}
\multiput(5,210)(20,0){11}{\line(1,0){10}}
\multiput(67.5,160)(20,0){11}{\line(1,0){10}}
\multiput(67.5,50)(20,0){11}{\line(1,0){10}}
\multiput(220,315)(0,-20){11}{\line(0,-1){10}}
\multiput(282.5,265)(0,-20){11}{\line(0,-1){10}}
\multiput(110,315)(0,-20){11}{\line(0,-1){10}}
\multiput(172.5,265)(0,-20){11}{\line(0,-1){10}}
\multiput(282.5,265)(0,-20){11}{\line(0,-1){10}}
\multiput(220,210)(25,-20){5}{\line(5,-4){20}}
\multiput(110,210)(25,-20){5}{\line(5,-4){20}}
\multiput(220,100)(25,-20){5}{\line(5,-4){20}}
\multiput(110,100)(25,-20){5}{\line(5,-4){20}}
\multiput(0,100)(110,0){3}{\circle*{10}}
\multiput(0,210)(110,0){3}{\circle*{10}}
\multiput(0,320)(110,0){3}{\circle*{10}}
\multiput(62.5,50)(110,0){3}{\circle*{10}}
\multiput(62.5,160)(110,0){3}{\circle*{10}}
\multiput(62.5,270)(110,0){3}{\circle*{10}}
\multiput(125,0)(110,0){3}{\circle*{10}}
\multiput(125,110)(110,0){3}{\circle*{10}}
\multiput(125,220)(110,0){3}{\circle*{10}}
\multiput(319,40)(0,1){3}{\line(1,0){100}}
\multiput(319,40)(0,1){3}{\line(2,1){20}}
\multiput(319,40)(0,1){3}{\line(2,-1){20}}
\multiput(345,92)(0,1){3}{\line(1,0){40}}
\multiput(340,92)(-15,0){6}{\line(-1,0){10}}
\multiput(340,93)(-15,0){6}{\line(-1,0){10}}
\multiput(340,94)(-15,0){6}{\line(-1,0){10}}
\multiput(255,92)(0,1){3}{\line(2,1){20}}
\multiput(255,92)(0,1){3}{\line(2,-1){20}}
\multiput(319,160)(0,1){3}{\line(1,0){100}}
\multiput(319,160)(0,1){3}{\line(2,1){20}}
\multiput(319,160)(0,1){3}{\line(2,-1){20}}
\multiput(210,290)(0,1){3}{\line(1,0){100}}
\multiput(210,290)(0,1){3}{\line(2,1){20}}
\multiput(210,290)(0,1){3}{\line(2,-1){20}}
\multiput(30,270)(0,1){3}{\line(-1,0){100}}
\multiput(30,270)(0,1){3}{\line(-2,1){20}}
\multiput(30,270)(0,1){3}{\line(-2,-1){20}}
\multiput(30,130)(0,1){3}{\line(-1,0){100}}
\multiput(30,130)(0,1){3}{\line(-2,1){20}}
\multiput(30,130)(0,1){3}{\line(-2,-1){20}}
\multiput(90,180)(0,1){3}{\line(-1,0){120}}
\multiput(90,180)(0,1){3}{\line(-2,1){20}}
\multiput(90,180)(0,1){3}{\line(-2,-1){20}}
\multiput(90,70)(0,1){3}{\line(-1,0){120}}
\multiput(90,70)(0,1){3}{\line(-2,1){20}}
\multiput(90,70)(0,1){3}{\line(-2,-1){20}}
\put(430,150){\Large $W_f$}\put(430,30){\Large $W_d$}
\put(396,82){\Large $W_h$}\put(321,280){\Large $W_b$}
\put(-110,120){\Large $W_c$}\put(-110,260){\Large $W_g$}
\put(-70,170){\Large $W_a$}\put(-70,60){\Large $W_e$}
\end{picture}}
\put(200,0){{\bf Fig. 6.1} A composite weight ${\cal W}$.}
\end{picture}

Let us demand that four weights ${\cal W}$, ${\cal W'}$, ${\cal W''}$ and
${\cal W'''}$ should satisfy to usual tetrahedron equations of mixed type:
\begin{equation}
{\cal W\> W'\> W''\> W'''}={\cal W'''\> W''\> W'\> W}. \label{two1}
\end{equation}
So constructing from ${\cal W}$ and ${\cal W'}$ two-layer transfer matrices
$T({\cal W})$ and $T({\cal W'})$ in a usual way we come to a commuting
relation:
\begin{equation}
[T({\cal W}),T({\cal W'})]=0               \label{two1a}
\end{equation}

Then one can show that it is equivalent to the validity of the following
16 equations of type (\ref{2}):
\begin{eqnarray}
&W_h W'_c W''_e W'''_a=\ov W_h\ov  W'_c\ov  W''_e\ov  W'''_a,\quad
\ov W_a \ov W'_f \ov W''_b \ov W'''_h= W_a W'_f W''_b W'''_h,&\nonumber\\
&W_c \ov W'_c W''_c W'''_g= \ov W_c W'_c \ov W''_c \ov W'''_g,\quad
\ov W_f W'_f \ov W''_f \ov W'''_d= W_f \ov W'_f W''_f W'''_d,&\nonumber\\
&W_b W'_g W''_a \ov W'''_a= \ov W_b \ov W'_g \ov W''_a W'''_a,\quad
\ov W_e \ov W'_d \ov W''_h W'''_h= W_e W'_d W''_h \ov W'''_h,&\nonumber\\
&W_d W'_e \ov W''_e W'''_e= \ov W_d \ov W'_e W''_e \ov W'''_e,\quad
\ov W_g \ov W'_b W''_b \ov W'''_b= W_g W'_b \ov W''_b W'''_b,\label{two2}\\
&\ov W_h W'_h W''_d W'''_f= W_h \ov W'_h \ov W''_d \ov W'''_f,\quad
W_a \ov W'_a \ov W''_g \ov W'''_c= \ov W_a W'_a W''_g W'''_c,&\nonumber\\
&W_e \ov W'_e \ov W''_c W'''_c= \ov W_e W'_e W''_c \ov W'''_c,\quad
\ov W_b W'_b W''_f \ov W'''_f= W_b \ov W'_b \ov W''_f W'''_f,&\nonumber\\
&W_f W'_a \ov W''_a \ov W'''_e= \ov W_f \ov W'_a W''_a W'''_e,\quad
\ov W_c \ov W'_h W''_h W'''_b= W_c W'_h \ov W''_h \ov W'''_b,&\nonumber\\
&W_g \ov W'_g W''_g \ov W'''_g= \ov W_g W'_g \ov W''_g W'''_g,\quad
\ov W_d W'_d \ov W''_d W'''_d= W_d \ov W'_d W''_d \ov W'''_d.&\nonumber
\end{eqnarray}
An arrangement of spin variables in relations (\ref{two1}) is the same as
in the modified tetrahedron equations (\ref{2}). Note that we must introduce
a set of additional weights $\ov W_a,\ldots,\ov W_h$ which appear only
in intermediate steps when proving relations (\ref{two1}) and cancel
after 16-fold application of relations (\ref{two2}).

Not let us analyze relations (\ref{two2}). The simplest way to satisfy
them is to put $W_g=W_e=W_f=W_h=W$ and $W_d=W_b=W_c=W_a=\ov W$.
In this case we must demand that $W$ and $\ov W$ should satisfy
relations (\ref{2}) and its dual version. This is the case considered
in paper \cite{NewTw} where it was shown that then we have only two solutions:
namely the Bazhanov-Baxter model and the ``static limit'' elliptic model.
So we should try to avoid this situation.

But using results of the Section 3 we can directly analyze solutions of system
(\ref{two2}). To simplify our discussion we will omit bellow all gauge
factors before curly brackets entering in formulae for weight function
(\ref{12}) and relations for corresponding variables $u,v,\l,\xi,\mu,\nu$.
They can be easily restored by using  relations like (\ref{32}-\ref{34}).

In Section 3 we have seen that all relations containing the coordinates
$\ov x$ follow (up to some roots of unity) from relations (\ref{24},\ref{35}).
So we will analyze only those constraints which contain
variables ${x_a}_i,{x_a}_{ij},\ldots,{x_h}_i,{x_h}_{ij}$.
Introducing notations like (\ref{20}-\ref{21}) for the rapidity variables
for the letters: $a,\ldots,h$ we come to the following relations:
\begin{eqnarray}
&{j_c}_1={i_g}_1,\quad{j_h}_1={i_b}_1,\quad{j_e}_1={i_a}_1,\quad
{j_d}_1={i_f}_1,\quad{j_b}_2={i_g}_2,\quad{j_f}_2={i_a}_2,&\nonumber\\
&{j_h}_2={i_c}_2,\quad{j_d}_2={i_e}_2,\quad{j_a}_3={i_g}_3,\quad
{j_f}_3={i_b}_3,\quad{j_e}_3={i_c}_3,\quad{j_d}_3={i_h}_3,&\label{two3}
\end{eqnarray}
\begin{equation}
a_g=a_e=a_f=a_h=a_d^{-1}=a_b^{-1}=a_c^{-1}=a_a^{-1}=a=a'=a''=a'''\label{two4}
\end{equation}
and the same for all $i',i'',i''',j',j'',j'''$.

Relations (\ref{two4}) show that modulus parameters $a_i$ associated with
elementary cubes of the lattice have a checkerboard structure in all
directions.

We also  have 24 relations like (\ref{24}):
\begin{equation}
{i_g}_1={i_g}_2''',\quad {i_g}_2={i_g}_2',\quad {i_g}_3={i_g}_2'',\quad
{i_g}_1'={i_g}_3''',\quad
{i_g}_3'={i_g}_3'',\quad {i_g}_1''={i_g}_1''',   \label{two5}
\end{equation}
\begin{equation}
{j_d}_1={j_d}_2''',\quad {j_d}_2={j_d}_2',\quad {j_d}_3={j_d}_2'',\quad
{j_d}_1'={j_d}_3''',\quad
{j_d}_3'={j_d}_3'',\quad {j_d}_1''={j_d}_1''',   \label{two6}
\end{equation}
and
\begin{equation}
{i_c}_2={i_c}_2',\quad {i_c}_3={i_c}_2'',\quad {i_c}_3'={i_c}_3'',\quad
{i_b}_3'={i_b}_3'',\quad
{i_b}_1'={i_b}_3''',\quad {i_b}_1''={i_b}_1''',   \label{two7}
\end{equation}
\begin{equation}
{i_a}_2={i_a}_2',\quad {i_a}_1''={i_a}_1''',\quad {i_a}_1'={i_c}_3''',\quad
{i_a}_1={i_c}_2''',\quad
{i_b}_1={i_a}_2''',\quad {i_b}_3={i_a}_2''.   \label{two8}
\end{equation}
At last, we have 16 tetrahedron constraints like (\ref{35}):
\begin{eqnarray}
&{\di {x_c}'_{58}\over\di {x_h}_{13}}
{\di {x_e}''_5\over\di {x_e}''_{58}}
{\di {x_a}'''_{24}\over\di {x_a}'''_2}=\w,\quad
{\di {x_a}_{24}\over\di {x_a}_{2}}
{\di {x_f}'_6\over\di {x_f}'_{67}}
{\di {x_b}''_{67}\over\di {x_h}'''_{13}}=1,\quad
{\di {x_f}_6\over\di {x_f}_{67}}
{\di {x_f}'_{67}\over\di {x_f}'_6}
{\di {x_f}''_6\over\di {x_f}''_{67}}
{\di {x_d}'''_{13}\over\di {x_d}'''_1}=1,&\nonumber\\
&\w{\di {x_b}_{67}\over\di {x_g}'_{24}}
{\di {x_a}''_{24}\over\di {x_a}''_2}
{\di {x_a}'''_2\over\di {x_a}'''_{24}}=1,\quad
{\di {x_e}_5\over\di {x_e}_{58}}
{\di {x_d}'_{13}\over\di {x_d}'_1}
{\di {x_h}'''_{13}\over\di {x_h}''_{13}}=1,\quad
{\di {x_d}_{13}\over\di {x_d}_1}
{\di {x_e}'_5\over\di {x_e}'_{58}}
{\di {x_e}''_{58}\over\di {x_e}''_5}
{\di {x_e}'''_5\over\di {x_e}'''_{58}}=1,&\nonumber\\
&{\di {x_h}_{13}\over\di {x_h}'_{13}}
{\di {x_d}''_{13}\over\di {x_d}''_1}
{\di {x_f}'''_6\over\di {x_f}'''_{67}}=1,\quad
{\di {x_e}_2\over\di {x_e}_{58}}
{\di {x_e}'_{58}\over\di {x_e}'_2}
{\di {x_c}'''_{58}\over\di {x_c}''_{58}}=1,\quad
{\di {x_b}_{67}\over\di {x_b}'_{67}}
{\di {x_f}''_{67}\over\di {x_f}''_2}
{\di {x_f}'''_2\over\di {x_f}'''_{67}}=1,&\label{two9}\\
&\w{\di {x_f}_{67}\over\di {x_f}_2}
{\di {x_a}'_2\over\di {x_a}'_{24}}
{\di {x_a}''_{24}\over\di {x_a}''_2}
{\di {x_e}'''_2\over\di {x_e}'''_{58}}=1,\quad
{\di {x_a}_2\over\di {x_a}_{24}}
{\di {x_a}'_{24}\over\di {x_a}'_2}
{\di {x_c}'''_{58}\over\di {x_g}''_{24}}=1,\quad
{\di {x_d}_{13}\over\di {x_d}_{1}}
{\di {x_d}'_1\over\di {x_d}'_{13}}
{\di {x_d}''_{13}\over\di {x_d}''_1}
{\di {x_d}'''_1\over\di {x_d}'''_{13}}=1,&\nonumber\\
&{\di {x_c}_{58}\over\di {x_c}'_{58}}
{\di {x_c}''_{58}\over\di {x_g}'''_{24}}=1,\quad
\w{\di {x_b}'_{67}\over\di {x_g}_{24}}
{\di {x_b}'''_{67}\over\di {x_b}''_{67}}=1,\quad
\w{\di {x_h}'_{13}\over\di {x_c}_{58}}
{\di {x_b}'''_{67}\over\di {x_h}''_{13}}=1,\quad
{\di {x_g}_{24}\over\di {x_g}'_{24}}
{\di {x_g}''_{24}\over\di {x_g}'''_{24}}=1.&\nonumber
\end{eqnarray}
Since relations (\ref{two3}-\ref{two4}) leave 13 degrees of freedom
in each of the composite weights (12 ``spectral'' and one modulus parameters),
we have 40 relations (\ref{5}-\ref{9}) on $12 \times 4+1=49$ variables.

One can show that all other relations constrained $x$ and $\ov x$ are
satisfied automatically up to some roots of unity provided that
relations (\ref{two3}-\ref{two9}) are valid.
The proof of this fact is  quite tedious and we will not give it
here. But we must be sure that there exists a consistent choice
of degrees of $\w$ in all parameters so that all 16 equations (\ref{two2})
will be valid. Below we will give a simplified solution of
system (\ref{two3}-\ref{two9}) and the whole system of equations (\ref{two2}).
It shows we should find a general solution only for relations
(\ref{two3}-\ref{two9}) and then all degrees of $\w$ in parameters
$x$ and $\ov x$ can be restored by a continuous deformation of the
parameters for the simplified solution.

Now let us turn to system (\ref{two3}-\ref{two9}). Unfortunately, up to
now we did not succeed in obtaining a complete understanding of the underlying
geometrical picture (if any).
If we put $a=1$ then we come to the inhomogeneous variant of the $N$-state
generalization of the Zamolodchikov model proposed by Bazhanov and Baxter.
In this case we have a natural geometrical picture.
Each elementary weight is described by the three angles
between the three planes. Then a two-layer composite cube is defined by
six intersecting planes and their mutual orientation depends on
nine angle variables. All system (\ref{two3}-\ref{two9}) has 13
degrees of freedom and therefore we have a four-parametric commuting
family of two-layer transfer-matrices.

Unfortunately, the case $a\ne1$ looks much more complicated due to the absence
of a geometrical analogy. Nevertheless, we have analyzed it quite
carefully. The results can be formulated as follows:
system (\ref{two3}-\ref{two9}) has 12 degrees of freedom (11 ``angle''
parameters and one modulus). A composite two-layer cube depends from
only from eight independent ``angles'' and modulus. So we have an
additional constraint among nine ``angles'' and this constraint will
disappear in the limit $a\to 1$. We also obtain only
a three-parametric family of commuting two-layer transfer-matrices.
We will not write here these formulae since they are rather cumbersome
and therefore practically useless.
Nevertheless, we are sure that there should exist a homogeneous solution
to the  system (\ref{two3}-\ref{two9}),so that
all weights $W_a,\ldots,W_h$ depend on the same three spectral
parameters and modulus) which will generalize the model of Bazhanov-Baxter
and ``static limit'' elliptic model of \cite{NewTw}.

Now we will consider another limit case of system (\ref{two3}-\ref{two9}).
Demand that all parameters $x_a,\ldots,x_h$ are expressed in terms of
$x_g$ by the following formulae:
\begin{eqnarray}
&{x_a}_1={\w/ {x_g}_1},\quad
{x_a}_2={\w/ {x_g}_2},\quad
{x_a}_5={\w^2/ {x_g}_5},\quad
{x_a}_6={1/ {x_g}_6},&\label{two10}\\
&{x_a}_{13}=\w^{1/2}{\di{x_g}_{13}\over\di {x_g}_1},\>
{x_a}_{24}=\w^{-1/2}{\di{x_g}_{24}\over\di {x_g}_2},\>
{x_a}_{58}=\w^{1/2}{\di{x_g}_{58}\over\di {x_g}_5},\>
{x_a}_{67}=\w^{-1/2}{\di{x_g}_{67}\over\di {x_g}_6},&\nonumber
\end{eqnarray}
\begin{eqnarray}
&{x_e}_1={1/ {x_g}_6},\quad
{x_e}_2={\w/ {x_g}_5},\quad
{x_e}_5={\w/ {x_g}_2},\quad
{x_e}_6={1/ {x_g}_1},&\label{two11}\\
&{x_e}_{13}=\w^{1/2}{\di{x_g}_{67}\over\di {x_g}_6},\>
{x_e}_{24}=\w^{1/2}{\di{x_g}_{58}\over\di {x_g}_5},\>
{x_e}_{58}=\w^{1/2}{\di{x_g}_{24}\over\di {x_g}_2},\>
{x_e}_{67}=\w^{-1/2}{\di{x_g}_{13}\over\di {x_g}_1},&\nonumber
\end{eqnarray}
\begin{eqnarray}
&{x_f}_1={\w/ {x_g}_5},\quad
{x_f}_2={1/ {x_g}_6},\quad
{x_f}_5={\w/ {x_g}_1},\quad
{x_f}_6={1/ {x_g}_2},&\label{two12}\\
&{x_f}_{13}=\w^{1/2}{\di{x_g}_{58}\over\di {x_g}_5},\>
{x_f}_{24}=\w^{1/2}{\di{x_g}_{67}\over\di {x_g}_6},\>
{x_f}_{58}=\w^{1/2}{\di{x_g}_{13}\over\di {x_g}_1},\>
{x_f}_{67}=\w^{-1/2}{\di{x_g}_{24}\over\di {x_g}_2},&\nonumber
\end{eqnarray}
\begin{eqnarray}
&{x_d}_1={1/ {x_g}_2},\quad
{x_d}_2={1/ {x_g}_1},\quad
{x_d}_5={1/ {x_g}_6},\quad
{x_d}_6={1/ {x_g}_5},&\label{two13}\\
&{x_d}_{13}=\w^{-1/2}{\di{x_g}_{24}\over\di {x_g}_2},\>
{x_d}_{24}=\w^{1/2}{\di{x_g}_{13}\over\di {x_g}_1},\>
{x_d}_{58}=\w^{1/2}{\di{x_g}_{67}\over\di {x_g}_6},\>
{x_d}_{67}=\w^{-1/2}{\di{x_g}_{58}\over\di {x_g}_5},&\nonumber
\end{eqnarray}
\begin{eqnarray}
&{x_b}_1={{x_g}_5/\w},\quad{x_b}_2={{x_g}_6},\quad
{x_b}_5={{x_g}_1},\quad{x_b}_6={{x_g}_2/\w},&\label{two14}\\
&{x_b}_{13}={x_g}_{58}/\w,\quad{x_b}_{24}=\w{x_g}_{67},\quad
{x_b}_{58}={x_g}_{13},\quad{x_b}_{67}={x_g}_{24}/\w,&\nonumber
\end{eqnarray}
\begin{eqnarray}
&{x_c}_1={{x_g}_6},\quad{x_c}_2={{x_g}_5/\w},\quad
{x_c}_5={{x_g}_2},\quad{x_c}_6={{x_g}_1/\w},&\label{two15}\\
&{x_c}_{13}={x_g}_{67},\quad{x_c}_{24}={x_g}_{58},\quad
{x_c}_{58}={x_g}_{24},\quad{x_c}_{67}={x_g}_{13}/\w,&\nonumber
\end{eqnarray}
\begin{eqnarray}
&{x_h}_1={{x_g}_2/\w},\quad{x_h}_2={{x_g}_1/\w},\quad
{x_h}_5={{x_g}_6},\quad{x_h}_6={{x_g}_5/\w^2},&\label{two16}\\
&{x_h}_{13}={x_g}_{24}/\w^2,\quad{x_h}_{24}={x_g}_{13},\quad
{x_h}_{58}={x_g}_{67},\quad{x_h}_{67}={x_g}_{58}/\w^2&\nonumber
\end{eqnarray}
and same for  $x'$, $x''$, $x'''$ and $\ov x$  $\ov x'$,
$\ov x''$, $\ov x'''$. Then all
relations (\ref{two3}-\ref{two9}) will be satisfied automatically
as a consequence of the relations for $x_g$ (\ref{two5}) and the tetrahedron
constraints for $x_g$ (see (\ref{35})). Moreover, all relations
containing $\ov x$ will be satisfied also. So we can use formulae
(\ref{10}-\ref{16}) in order to determine all $\w$ multipliers.
Now we will show that transformations (\ref{two10}-\ref{16})
have a natural interpretation.

First consider a transformation of $x$ given by (\ref{two13}). Let us denote
a weight function from this  set of parameters as $W^{(i)}$.
Then one can show that $W$ and $W^{(i)}$ satisfy to the following
inversion relation:
\begin{equation}
\sum_{\s\in Z_N}W(a|efg|bcd|\s)W^{(i)}(\s|bcd|efg|h)=\Phi(x)\delta_{a,h},
\label{two17}
\end{equation}
where $\Phi(x)$ is some scalar factor which can be calculated.

Define also the following automorphism $\theta$ on $x's$ as:
\begin{eqnarray}
&\theta(x_1,x_2,x_5,x_6,x_{13},x_{24},x_{58},x_{67})=\nonumber\\&(x_1/\w,
x_2/\w,x_5/\w,x_6/\w,x_{13}/\w^2,x_{24},x_{58}/\w,x_{67}/\w).&\label{two18}
\end{eqnarray}
Then all relations (\ref{two10}-\ref{two16}) can be rewritten
in the following compact form:
\begin{eqnarray}
&W_f=W_g^{\rho^2},\quad W_e=W_g^{\tau\rho^2\tau},\quad
W_h=W_g^{\theta\tau\rho^2\tau\rho^2}&,\nonumber\\
&W_a=W_h^{(i)},\quad W_b=W_e^{(i)},\quad W_c=W_f^{(i)},\quad W_d=W_g^{(i)},&
                                \label{two19}
\end{eqnarray}
where the action of the transformations $\tau$ and $\rho$ on $x's$
was defined in Section 5 and the action of their compositions can be
easily calculated.
We call the choice of weights in the form (\ref{two19}) as the ``inverse''
variant of two-layer model.

Note that we were forced to introduce automorphism $\theta$ in order
to satisfy all the systems of relations for $x's$ and $\ov x's$ coming from
(\ref{two2}). This is a very overdetermined system of scalar equations
and it is remarkable that we have any solution at all.

\section{Conclusion}
In this paper we have found a new solution to the modified tetrahedron
equations in the framework of the so called ``Body-Centered-Cube'' ansatz.
The weight functions depend on three angle-like parameter $\theta_{i}$
and one elliptic modulus $k$. We have also constructed the elliptic
parameterization of the weight function. Note that in the limit
$k \rightarrow 0$ we obtain the Baxter-Bazhanov model . If the
``static limit'' condition $\theta_{1}+\theta_{2}+\theta_{3} = 2K$
with $K$ being the full elliptic integral is satisfied than we obtain
the model proposed in \cite{Man,NewTw}.
As was mentioned above we were forced to
achieve the integrability of the model without the validity of ``dual''
version of MTE to be able to generalize the Baxter-Bazhanov model and
not to restrict ourself onto ``static-limit'' condition.
The constructed integrable two-layer model has the spin variables
in the cube vertices,the middle points of the cube edges and the centers of
the cube faces as was shown in fig. $6.1$ . So,this model is of mixed type.
The weight function consist of eight weights
$W_{a},W_{b},...W_{h}$ which must satisfy the system of the
sixteen sets of equations (6.3). The solution to these equations appeared to
have twelve free parameters (including the modulus $k$) but the
formulae are rather cumbersome and we do not write them. We have
also considered one particular case when all four pairs of the opposite
weights (a-h,b-e,c-f,d-g) are connected with each other by the inversion.
Unfortunately, we did not succeed to find a geometrical picture like
in the case of the Baxter-Bazhanov model. Nevertheless some of the
formulae presented in Appendix resemble the formulae for
the spherical triangle taking place for the angles $\theta_{1},\theta_{2},
\theta_{3}$ in the Baxter-Bazhanov model.
 So, we hope that some appropriate geometrical picture does exist.
We also hope that besides the ``inverse'' model
some another two-layer ``homogeneous'' model which might  have
a rich physical content can be found. We expect to do this elsewhere.

\noindent
{\bf Acknowledgement}

\noindent
This work is partially supported by the International Scentific
Fund (INTAS), Grant No. RMM000. The authors would like to thank
A. Shiekh (ICTP, Trieste) for his valuable advises.

\section{Appendix}
The formulae for ``modified spherical triangle'' are more aesthetic if we
introduce a parameterization for $T_{1}$,$T_{2}$,$T_{3}$
in terms of the elliptic
functions of the modulus $k = (1-m)/(1+m)$.
Besides, some of the formulae bellow resemble corresponding formulae
in the spherical trigonometry.
We shall use the ordinary Jacobi's elliptic
functions $\sn$, $\cn$ and $\dn$, theta -- and eta -- functions
$H$ , $\Theta $ .
Let us also introduce  the following notations:
\begin{equation}
\di\tn(u,k) = k'{\di\sn(u,k)\over\di\cn(u,k)\dn(u,k)}\label{a1}
\end{equation}
and
\begin{eqnarray}
&\di{\tn}_L(u,k) = \tn((1+k)u,{\di 2k^{1/2}\over\di 1+k}),\quad
\ov{\tn}_L(u,k) = \tn((1-k)u,{\di 2ik^{1/2}\over\di 1-k})&\nonumber\\
&\di{\tn}_L(u,k)\ov{\tn}_L(u,k)=\tn^2(u,k).&\label{a2}
\end{eqnarray}
The last two functions represent the Landen's transformation
$\tau\rightarrow\tau/2$ and $\tau\rightarrow\tau/2+1$.
The parameters $k$ and $m$ we regard to be fixed forever and so we shall
omit them everywhere. Also we shall use the indexes $p,q,r$ for $1,2,3$

Defining $\t_r$ so as
\begin{equation}
T_r = \tn(\t_r/2),\label{a3}
\end{equation}
then we obtain
\begin{equation}
S_r = \sn(\t_r),\quad C_r = \cn(\t_r), \quad D_r = \dn(\t_r).
\label{a4}
\end{equation}
Having got the angles $\t_r$, define the spherical excesses as usual:
\begin{equation}
\di\a_0 = {\di\t_1+\t_2+\t_3\over\ov 2} - {\cal K},\quad
\a_r = \t_r-\a_0,\label{a5}
\end{equation}
where $\cal K$ is the complete elliptic integral of the first kind of the
modulus $k$. It can be shown that the following expressions are valid:
\begin{eqnarray}
&\di \sin^2{a_0/2} = {\di H(\a_0)H(\a_1)H(\a_2)H(\a_3)\over
\di\T(0)\T(\t_1)\T(\t_2)\T(\t_3)},&\nonumber\\
&\di \cos^2{a_0/2} = {\di\T(\a_0)\T(\a_1)\T(\a_2)\T(\a_3)\over
\di\T(0)\T(\t_1)\T(\t_2)\T(\t_3)},&\nonumber\\
&\di\tan{\di a_0\over\di 2} = k\sqrt{\sn(\a_0)\sn(\a_1)\sn(\a_2)\sn(\a_3)};&
\label{a6}
\end{eqnarray}
\begin{eqnarray}
&\di \sin^2{a_r/2} = {\di H(\a_0)H(\a_r)\T(\a_p)\T(\a_q)\over
\di\T(0) \T(\t_r) H(\t_p) H(\t_q)},&\nonumber\\
&\di \cos^2{a_r/2} = {\di\T(\a_0)\T(\a_r) H(\a_p) H(\a_q)\over
\di\T(0)\T(\t_r) H(\t_p) H(\t_q)},&\nonumber\\
&\di\tan{\di a_r\over\di 2} = \sqrt{\di\sn(\a_0)\sn(\a_r)\over\di
\sn(\a_p)\sn(\a_q)}.&\label{a7}
\end{eqnarray}
Similarly one can  obtain for the  linear excesses:
\begin{eqnarray}
\di\tan^2{\di\b_0\over\di 2} = {\di\tn_L(\a_1/2)\tn_L(\a_2/2)\tn_L(\a_3/2)
\over\di\tn_L(\a_0/2)},&\nonumber\\
\di\tan^2{\di\b_r\over\di 2} = {\di\tn_L(\a_0/2)\tn_L(\a_p/2)\tn_L(\a_q/2)
\over\di\tn_L(\a_r/2)}.&\label{a8}
\end{eqnarray}
The formulae for the  $\ov\b$ can be obtained from (\ref{a8}) by
replacing all the ${\tn}_L$ by $\ov{\tn}_L$ (it is the common rule: replacing
$k$ by $-k$ is equivalent to the modular transformation $\tau$ to $\tau +
2$).  We complete our list of the formulae by
\begin{eqnarray}
&\di{\di\sin\b_r\over\di\sin\b_0} = \tn_L(\t_q/2)\tn_L(\t_p/2),&\nonumber\\
&\di{\di\sin\b_p\sin\b_q\over\di\sin\b_0\sin\b_r} =
\tn_L^2(\t_r/2),\quad{\di\sin\b_p\sin\b_q\over\di\sin\ov\b_0\sin\ov\b_r} =
m\tn^2(\t_r/2).&\label{a9}
\end{eqnarray}

Now let us give the explicit form of the Boltzmann weight function for the
case $N = 2$. The calculations resembles ones from the Appendix
of \cite{NewTw} and we give here only the answer.
In our previous sections all the spins belong to $Z_N$
and take the values $0,1,...,N-1$. For the case $N=2$
it is convenient to deal with multiplicative spins $(-1)^a = \pm 1$
instead of $a=0,1$. Further we will imply the multiplicative spins as the
arguments of the Boltzmann weight functions.

Introduce for shortness some conventional notations:
\begin{equation}
\di s_i = \sqrt{\sn(\a_i/2)},\quad c_i = \sqrt{\cd(\a_i/2)},
\quad i = 0,1,2,3;\label{a10}
\end{equation}
and
\begin{eqnarray}
&\di P_0 = c_0c_1c_2c_3,\quad P_r = c_0c_rs_ps_q;&\nonumber\\
&\di Q_0 = s_0s_1s_2s_3,\quad Q_r = s_0s_rc_pc_q;&\nonumber\\
&\di R_i = s_ic_i,\quad S_i = k{\di s_0c_0s_1c_1s_2c_2s_3c_3\over\di s_ic_i},
\quad i=0,1,2,3.&\label{a11}
\end{eqnarray}
Then for the weight $W$ given by (\ref{12}) there exists the following
identity:
\begin{equation}
\di W = {\di 1\over\di P_0-Q_0} W_{SPQR}\label{a12}
\end{equation}
and $W_{SPQR}$ is given by the following table:

\vskip 0.5cm

\begin{center}
\begin{tabular}{||c|c|c||l||}
\hline
&&&\\
$abeh$&$acfh$&$adgh$&$ W_{SPQR}(a|efg|bcd|h)$\\ &&&\\ \hline
$+$&$+$&$+$&$P_0-abcdQ_0$\\
$-$&$+$&$+$&$R_1+abcdS_1$\\
$+$&$-$&$+$&$R_2+abcdS_2$\\
$+$&$+$&$-$&$R_3+abcdS_3$\\
$+$&$-$&$-$&$abP_1+cdQ_1$\\
$-$&$+$&$-$&$acP_2+bdQ_2$\\
$-$&$-$&$+$&$adP_3+bcQ_3$\\
$-$&$-$&$-$&$R_0+abcdS_0$\\ \hline
\end{tabular}
\end{center}
\vskip 0.5cm

Note that for $W_{SPQR}$ all the symmetry properties are quite evident.

\end{document}